\documentstyle[12pt]{article}

\newcommand{\be}{\begin{eqnarray}}
\newcommand{\ee}{\end{eqnarray}}
\newcommand{\balpha}{\mbox{\boldmath $\alpha$}}
\newcommand{\bgamma}{\mbox{\boldmath $\gamma$}}
\newcommand{\bsigma}{\mbox{\boldmath $\sigma$}}
\begin{document}

\title{\bf Quasi-exact Solvability of \\ Dirac-Pauli Equation
\\ and generalized Dirac oscillators }

\author{Choon-Lin Ho$^1$ and Pinaki Roy$^2$}
\date
{\small \sl $^1$Department of Physics, Tamkang University, Tamsui
25137, Taiwan, R.O.C. \\ $^2$Physics and Applied Mathematics Unit,
Indian Statistical Institute, Calcutta 700035, India}

\maketitle

\begin{abstract}
In this paper we demonstrate that neutral Dirac particles in
external electric fields, which are equivalent to generalized
Dirac oscillators, are physical examples of quasi-exactly solvable
systems. Electric field configurations permitting quasi-exact
solvability of the system based on the $sl(2)$ symmetry are
discussed separately in spherical, cylindrical, and Cartesian
coordinates. Some exactly solvable field configurations are also
exhibited.
\end{abstract}

\vskip 3cm \noindent{Dec 5, 2003}

\newpage
\centerline{\mbox{\large {\bf I.  Introduction}}}

Search for exact solutions of wave equations, whether non
relativistic or relativistic, has been there since the birth of
quantum mechanics. In the case of non relativistic wave equations,
i.e.  Schr{\"o}dinger equations,  the class of potentials for
which the complete spectrum or a part of the spectrum are known
has grown over the years. However in the case of relativistic wave
equations, in particular, the Dirac equation very few exactly
solvable electromagnetic field configurations like homogeneous
magnetic fields \cite{rabi}, homogeneous electrostatic fields
\cite{sauter}, constant parallel magnetic fields \cite{lam} etc.
are known.

It may be mentioned that while exact solvability is desirable, in
practice it is not always possible to determine the whole
spectrum. In non relativistic quantum mechanics a new class of
potentials which are intermediate to exactly solvable ones and non
solvable ones have been found recently. These are called
quasi-exactly solvable (QES) problems for which it is possible to
determine algebraically a part of the spectrum but not the whole
spectrum [\cite{TU}-\cite{Ush}]. Although there are exceptions,
usually a QES problem admits a certain underlying Lie algebraic
symmetry which is responsible for the quasi-exact solutions.  For
Dirac equations, it had been shown that Dirac equation with
Coulomb interaction supplemented by a linear radial potential
\cite{BK}, and planar Dirac equation with Coulomb and homogeneous
magnetic fields \cite{Ho} are QES systems. More recently, the
quasi-exact solvability of the  Pauli equation was exhibited
\cite{HoRoy}. This immediately implies that the Dirac equation
coupled minimally to a vector potential is QES, since the square
of the Dirac Hamiltonian in this form is proportional to the Pauli
equation, up to an additive constant, namely, the square of the
rest energy.

Interactions of spin-1/2 fermions with electromagnetic fields are
usually introduced in the Dirac equation following the minimal
coupling prescription. However, this prescription is not the only
possibility.  For instance, the interaction due to anomalous
magnetic moment is incorporated via the non-minimal coupling
procedure, the resulting equation being called the Dirac-Pauli
equation \cite{D-P}.  An interesting example is neutral fermions
interacting with electromagnetic fields. There are a number of
interesting phenomena involving non minimal coupling of neutral
fermions to external
 electric fields. We recall that, the celebrated Aharonov-Casher effect
\cite{AC-eff}, which has been observed experimentally
\cite{AC-exp} is an example of such an interaction. Another
interesting system involving neutral fermions is the Dirac
oscillator, which is of considerable interest  in quantum
chromodynamics \cite{Dirac-osc}. However, unlike the case with
minimal coupling, studies in exact solutions of the Dirac
equations with non-minimal couplings are rather scanty
\cite{OC,SV,Lin}, not to mention studies in quasi-exact solutions
of these systems.  Only recently, efforts have been directed to
exploring certain structure relating to such systems. For
instance, the  underlying supersymmetry (SUSY) of the
Aharonov-Casher system \cite{bruce}, as well as that of a system
of neutral fermions interacting with a central electric field were
pointed out \cite{Semenov}.

In the present paper we shall examine quasi-exact solvability of
Dirac equation coupled non-minimally to external electric fields.
In particular,  based on the underlying $sl(2)$ symmetry we shall
determine the forms of electric fields which give rise to QES
Dirac-Pauli equations.

The organization of the paper is as follows.  In Sect.II the
Dirac-Pauli equation is presented, and its equivalence with the
generalized Dirac oscillators demonstrated.  Sect.III discusses
the exact and quasi-exact solvability of the Dirac-Pauli equation
with electric fields in the spherical coordinates.  Sect.IV
and V are devoted to the same problems with electric fields in the
cylindrical and Cartesian coordinates, respectively.  Sect.VI
concludes the paper.

\vskip 1 truecm

\centerline{\mbox{\large{\bf II.  The Dirac-Pauli Equation for
neutral Dirac particles }}}

We consider the motion of a neutral fermion of spin-1/2 with mass
$m$ and an anomalous magnetic moment $\mu$, in an external
electromagnetic field described by the field strength
$F_{\mu\nu}$.   The fermion is described by a four-component
spinor  $\Psi$ which obeys the Dirac--Pauli equation \cite{D-P}
\be
(i\gamma^\mu\partial_\mu-\frac{1}{2}\mu
\sigma^{\mu\nu}F_{\mu\nu}-m)\Psi=0~,\label{DP}
 \ee
 where
$\gamma^\mu=(\gamma^0, {\bgamma})$ are the Dirac matrices
satisfying \be\{\gamma^\mu,\gamma^\nu\}=2g^{\mu\nu} \ee
 with
$g^{\mu\nu}={\rm diag}(1,-1,-1,-1)$, and
\be
\sigma^{\mu\nu}={i\over 2}[\gamma^\mu, \gamma^\nu]~. \ee
In terms
of the external electric field {\bf E} and magnetic field {\bf B},
one can rewrite the second term in eq.(\ref{DP}) as
\be
\frac{1}{2}\sigma^{\mu\nu}F_{\mu\nu}= i{\balpha}\cdot{\bf
E}-{\bf\Sigma}\cdot{\bf B}~,
 \ee
 where
${\balpha}=\gamma^0{\bgamma}$, and $\Sigma^k=\frac 12\epsilon
^{kij}\sigma^{ij}$.  Here $\epsilon^{kij}$ is the totally
anti-symmetric tensors with $\epsilon^{123}=1$. For
time-independent fields one may set
\be
\Psi(t,{\bf r}) =e^{-i{\cal E}t}\psi({\bf r})~, \ee
 and eq.(\ref{DP}) becomes
 \be  H\psi={\cal
E}\psi,\ee
 with the Hamiltonian $H$ being given by
\be
H={\balpha}\cdot{\bf p}+i\mu{\bgamma}\cdot{\bf E}-
\mu\beta{\bf\Sigma}\cdot{\bf B}+\beta m~,\ee
 where ${\bf p}=-i\nabla$ and $\beta=\gamma^0$.

In this paper we shall consider cases with only electric
fields described by the the Hamiltonian \be H={\balpha}\cdot{\bf
p}+i\mu{\bgamma}\cdot{\bf E} +\beta m~\label{H}~.\ee
We choose the Dirac matrices in the standard representation
\be
{\balpha} =
\left( \begin{array}{cc} 0 & {\bsigma}\\
{\bsigma} & 0
\end{array}\right)~,~~~~~
\beta= \left( \begin{array}{cc} 1 & 0\\ 0 & -1
\end{array}\right)~,
\ee where $\bsigma$ are the Pauli matrices. We also define
$\psi=(\chi, \varphi)^t$, where $t$ denotes transpose, and both
$\chi$ and $\varphi$  are two-component spinors. Then the
Dirac--Pauli equation becomes
\be
{\bsigma}\cdot({\bf p}-i\mu {\bf E})\chi &=&({\cal E}+m)\varphi~,\nonumber\\
 {\bsigma}\cdot({\bf p}+i\mu {\bf
E})\varphi &=&({\cal E}-m)\chi,\label{H1} \ee

In the next three sections, we will demonstrate the exact and
quasi-exact solvability of this set of equations in the spherical,
cylindrical, and Cartesian coordinates.

\vskip 0.8cm

 \centerline{\bf  Relation with the Dirac oscillator}

Here we show that the above system is equivalent to the
generalized Dirac oscillators.  Since its introduction by
Moshinsky and Szczepaniak in 1989, the original Dirac
oscillator has attracted much
attention in recent years \cite{Dirac-osc}.  It was introduced by
replacing the momentum $\bf p$ in the
field-free Dirac equation by ${\bf p} -im\omega\beta {\bf r}$,
where $m$ is the mass of the particle and $\omega$ is the
oscillation frequency.  The Hamiltonian of this system is thus
given by
\be
H={\balpha}\cdot ({\bf p}-im\omega\beta {\bf r}) +\beta m~.
\label{D-osc} \ee This system is exactly solvable.

Now if one compares eq.(\ref{D-osc}) with (\ref{H}), one sees
immediately that the Dirac oscillator is simply a special case of
neutral Dirac particle in external electric field, if one makes
the transformation
\be
\mu {\bf E} \to m\omega{\bf r}~. \ee That is, when the electric
field $\bf E$ is proportional to the position vector $\bf r$.
Hence, with a general $\bf r$-dependent electric field $\bf E(r)$,
the system described by the Hamiltonian (\ref{H}) can be viewed as
generalizations of the original Dirac oscillator.  In the
discussions below, we shall concentrate on the neutral Dirac
particles.  But all conclusions thus obtained are applicable
directly to the generalized Dirac oscillators.

\vskip 1 truecm

\centerline{\mbox{\large{\bf III.  Electric fields in spherical
coordinates}}}

First let us consider central electric field ${\bf E}=E_r
{\hat{\bf r}}$. In this case, one can choose a complete set of
observables to be $(H,{\bf J}^2,J_z,{\bf S}^2=3/4,K)$. Here $\bf
J$ is the total angular momentum ${\bf J=L+S}$, where $\bf L$ is
the orbital angular momentum, and ${\bf S}=\frac{1}{2}{\bf\Sigma}$
is the spin operator. The operator $K$ is defined as
$K=\gamma^0({\bf\Sigma}\cdot{\bf L}+1)$, which commutes with both
$H$ and {\bf J}. Explicitly, we have
\be
K&=&{\rm diag }\left({\hat k},-{\hat k}\right)~,\nonumber\\ {\hat
k}&=&\bsigma\cdot {\bf L} +1~. \ee
 The common eigenstates can be
written as \cite{Gross}
\be
\psi=\frac{1}{r} \left( \begin{array}{c} f_-(r) {\cal Y}^k_{jm_j}\\
if_+(r){\cal Y}^{-k}_{jm_j}
\end{array}\right)
\ee here ${\cal Y}^k_{jm_j}(\theta,\phi)$ are the spin harmonics
satisfying \be {\bf J}^2 {\cal Y}^k_{jm_j} &=& j(j+1){\cal
Y}^k_{jm_j}~,~~j=\frac{1}{2},\frac{3}{2},\ldots~~,
\\ J_z {\cal Y}^k_{jm_j} &=& m_j{\cal
Y}^k_{jm_j}~,~~~~~~~~|m_j|\leq j~~,
\\ {\hat k}{\cal Y}^k_{jm_j}&=& -k{\cal Y}^k_{jm_j}~,~~~~~~~~~
k=\pm(j+\frac{1}{2})~,  \ee and
\be
({\bsigma}\cdot {\hat {\bf r}}){\cal Y}^k_{jm_j}= -{\cal
Y}^{-k}_{jm_j}~, \ee  where $\hat{\bf r}$ is the unit radial
vector.
 Using the identity
\be
{\bsigma}\cdot {\bf p}=i({\bsigma}\cdot {\hat {\bf
r}})\left(-\partial_r+ \frac {1}{r}({\bsigma}\cdot {\bf
L})\right)~, \ee one gets
\be
{\bsigma}\cdot ({\bf p\pm i\mu {\bf E}})=i({\bsigma}\cdot {\hat {\bf
r}})\left(-\partial_r+
\frac {K-1}{r}\pm \mu E_r\right)~.
\ee
With these, eq.(\ref{H1}) reduces to
\be
\left(\frac{d}{dr} + \frac{k}{r} + \mu E_r\right)f_- &=&
\left({\cal E} + m\right)f_+~,\label{f-}\\ \left(-\frac{d}{dr} +
\frac{k}{r} + \mu E_r\right)f_+ &=& \left({\cal E} - m\right)f_-~.
\label{f+} \ee This shows that $f_-$ and $f_+$ forms a
one-dimensional SUSY pairs \cite{Semenov}. For if
we write the superpotential $W$ as
\be
W=\frac{k}{r} + \mu E_r~,
\label{W1}
\ee
 then eqs.(\ref{f-}) and (\ref{f+}) become
\be
A^-A^+f_-&=&\left({\cal E}^2 -m^2\right)f_-~,\\
A^+A^-f_+&=&\left({\cal E}^2 -m^2\right)f_+~, \ee with
\be
A^\pm\equiv \pm \frac{d}{dr}  +W~. \label{A} \ee Explicitly, the
above equations read
\be
\left(-\frac{d^2}{dr^2} + W^2 \mp W^\prime\right)f_\mp =
\left({\cal E}^2 - m^2\right)f_\mp~. \label{susy-1} \ee Here and
below the prime means differentiation with respect to the basic
variable.  Eq.(\ref{susy-1}) clearly exhibits the SUSY structure
of the system.  The operators acting on $f_\pm$ in
eq.(\ref{susy-1}) are said to be factorizable, i.e. as products of
$A^-$ and $A^+$. The ground state, with ${\cal E}^2=m^2$,  is
given by one of the following two sets of equations:
\be
A^+ f_-^{(0)}(r)&=& 0~~,~~~ f_+^{(0)}(r)=0~;\\ A^- f_+^{(0)}(r)&=&
0~~, ~~~f_-^{(0)}(r)=0~, \ee
depending on which solution is
normalizable. The solutions are generally given by
\be
f_\mp\propto r^{\mp k} \exp\left(\mp\int dr \mu E_r\right)~.
\ee

We now classify the forms of the electric field $E_r(r)$ which
allow exact and quasi-exact solutions. To be specific, we consider
the situation where $k<0$ and $\int dr \mu E_r >0$, so that
$f_-^{(0)}$ is normalizable, and $f_+^{(0)}=0$. The other
situation can be discussed similarly. In this case, eq.(\ref{W1})
becomes
\be
W=-\frac{|k|}{r} + \mu E_r~. \label{W2}
 \ee

We determine the forms of $E_r$ that give
exact/quasi-exact energy $\cal E$ and the corresponding function $f_-$.  The
corresponding function $f_+$ is obtained using eq.(\ref{f-}).

\vskip 0.8cm

 \centerline{{\bf A.  Exactly solvable cases}}

Comparing the forms of the superpotential $W$ in eq.(\ref{W2})
with Table~(4.1) in \cite{Cooper}, one concludes that there are three
forms of $E_r$ giving exact solutions of the problem :

~~~~~~~~~~i)   oscillator-like :  $\mu E_r(r)\propto r $~;

~~~~~~~~~~ii)  Coulomb potential-like :   $\mu E_r(r)\propto {\rm constant}
$~;

~~~~~~~~~~iii) zero field-like :  $\mu E_r(r)\propto 1/r $~.

Case (i) and (ii) had been considered in \cite{SV} and \cite{Lin}, and case
(iii) in \cite{SV}.

We mention here that the case with oscillator-like field, i.e. case (i), is
none other than the spherical Dirac oscillator.

\vskip 0.8cm

\centerline{{\bf B.  Quasi-exactly solvable cases}}

The form of the superpotential $W$ in eq.(\ref{W2}) fits into three
classes, namely, Classes VII, VIII and IX,
of $sl(2)$-based QES systems in \cite{Tur}.
Electric field configurations
permitting any number of solvable excited states in each of these classes can
be constructed according to the general procedure given in
\cite{HoRoy}.  The procedure makes use of the connection between quasi-exact
solvability and the SUSY structure (or, equivalently, the factorizability) of
the systems.   Below we describe the construction
of QES $E_r$ for the simplest cases in Class VII.  Other classes can be
considered similarly.

\vskip 0.8cm

\centerline{\bf  An outline of the method of construction}

Here we describe briefly the main ideas underlying the
construction of QES systems as given in \cite{HoRoy}.  As before,
we shall concentrate only on solution of the upper component
$f_-$, which is assumed to have a normalizable zero energy state.

Eq.(\ref{susy-1}) shows that $f_-$ satisfies the Schr\"odinger
equation $H_- f_-=\epsilon f_-$, with energy parameter $\epsilon
\equiv {\cal E}^2 - m^2 $, and
\be
H_-&=& A^-A^+ \nonumber\\
& =&-\frac{d^2}{dr^2} + V(r) ~,\label{SE} \ee
with
\be
V(r)=W(r)^2 -W^\prime(r)~. \label{V} \ee We shall look for $V(r)$
such that the system is QES.  According to the theory of QES
models, one first makes an imaginary gauge transformation on the
function $f_-$
\be
f_-(r)= \phi(r) e^{-g(r)}~, \label{f-1}
\end{eqnarray}
where $g(r)$ is called the gauge function.  The function $\phi(r)$
satisfies
\be
-\frac{d^2\phi(r)}{dr^2} + 2 g^\prime \frac{d\phi(r)}{dr} +
\left[V(r)+ g^{\prime\prime} - g^{\prime 2}\right]\phi
(r)=\epsilon\phi(r)~. \label{phi}
\end{eqnarray}
For physical systems which we are interested in, the phase factor
$\exp(-g(r))$ is responsible for the asymptotic behaviors of the
wave function so as to ensure normalizability. The function
$\phi(r)$ satisfies a Schr\"odinger equation with a gauge
transformed Hamiltonian
\be
H_G=-\frac{d^2}{dr^2} + 2W_0(r)\frac{d}{dr}  +\left[V(r)
+W_0^\prime - W_0^2\right]~, \label{HG} \ee where $W_0(r)=g^\prime
(r)$.  Now if $V(r)$ is such that the quantal system is QES, that
means the gauge transformed Hamiltonian $H_G$ can be written as a
quadratic combination of the generators $J^a$ of some Lie algebra
with a finite dimensional representation.  Within this finite
dimensional Hilbert space the Hamiltonian $H_G$ can be
diagonalized, and therefore a finite number of eigenstates are
solvable. For one-dimensional QES systems the most general Lie
algebra is $sl(2)$ (\cite{TU}-\cite{Ush}).  Hence if eq.(\ref{HG})
is QES then it can be expressed as
\be
H_G=\sum C_{ab}J^a J^b + \sum C_a J^a + {\rm constant}~,
\label{H-g}
\ee
where $C_{ab},~C_a$ are constant coefficients, and the $J^a$ are
the generators of the Lie algebra $sl(2)$ given by
\begin{eqnarray}
J^+ &=& z^2 \frac{d}{dz} - Nz~,\cr
J^0&=&z\frac{d}{dz}-\frac{N}{2}~,~~~~~~~~N=0,1,2\ldots\cr J^-&=&
\frac{d}{dz}~.
\end{eqnarray}
Here the variables $r$ and $z$ are related by $z=h(r)$, where
$h(\cdot)$ is some (explicit or implicit) function . The value
$j=N/2$ is called the weight of the differential representation of
$sl(2)$ algebra, and $N$ is the degree of the eigenfunctions $\phi$,
which are polynomials in a $(N+1)$-dimensional Hilbert space
\be
\phi=(z-z_1)(z-z_2)\cdots (z-z_N)~.
\label{phi-2}
\ee

The requirement in eq.(\ref{H-g}) fixes $V(r)$ and $W_0(r)$, and
$H_G$ will have an algebraic sector with $N+1$ eigenvalues and
eigenfunctions.  For definiteness, we shall denote the potential
$V$ admitting $N+1$ QES states by $V_N$.   From eqs.(\ref{f-1})
and (\ref{phi-2}), the function $f_-$ in this sector has the
general form
\be
f_-=(z-z_1)(z-z_2)\cdots (z-z_N)\exp\left(-\int^z W_0(r) dr\right)~,
\label{psi-1}
\end{eqnarray}
where $z_i$ ($i=1,2,\ldots,N$) are $N$ parameters that can be
determined by plugging eq.(\ref{psi-1}) into eq.(\ref{phi}).  The
algebraic equations  so obtained are called the Bethe ansatz
equations corresponding to the QES problem \cite{Ush,Ho}.  Now one
can rewrite eq.(\ref{psi-1}) as
\be
f_- =\exp\left(-\int^z W_N(r,\{z_i\}) dr\right)~,
\label{f2}
\ee
and
\be
W_N(r,\{z_i\}) = W_0(r) -  \sum_{i=1}^N
\frac{h^\prime(r)}{h(r)-z_i}~. \label{W}
\end{eqnarray}
There are $N+1$ possible functions $W_N (r,\{z_i\})$ for the $N+1$
sets of eigenfunctions $\phi$. Inserting eq.(\ref{f2}) into $H_-
f_-=\epsilon f_-$, one sees
that $W_N$ satisfies the Ricatti equation \cite{Shifman,Roy}
\begin{eqnarray}
W_N^2 - W_N^\prime = V_N - \epsilon_N~, \label{Ricatti}
\end{eqnarray}
where $\epsilon_N$ is the energy parameter corresponding to the
eigenfunction $f_-$ given in eq.(\ref{psi-1}) for a particular set
of $N$ parameters $\{z_i\}$.

From eqs.(\ref{SE}), (\ref{V}) and (\ref{Ricatti}) it is clear how
one should proceed to determine the electric fields so that the
Dirac-Pauli equation becomes QES based on $sl(2)$: one needs only
to determine the superpotentials $W(r)$ according to
eq.(\ref{Ricatti}) from the QES potentials $V(r)$  classified in
\cite{Tur}. This is easily done by observing that the
superpotential $W_0$ corresponding to $N=0$ is related to the
gauge function $g(r)$ associated with a particular class of QES
potential $V(r)$ by $g^\prime (r)=W_0 (r)$. Once $W_0$ is
obtained, the corresponding electric field $E^{(0)}$ is obtained
through eq.(\ref{W2}):
\be
\mu E^{(0)}_r=W_0 + |k|/r ~. \label{E-0} \ee
 This is the
required electric field that allows the weight zero ($j=N=0$)
state, i.e. the ground state, to be known in that class.  The more
interesting task is to obtain higher weight states (i.e. $j>0$),
which will include excited states.  For weight $j$ ($N=2j$)
states, this is achieved by forming the superpotential
$W_N(r,\{z_i\})$ according to eq.(\ref{W}). Of the $N+1$ possible
sets of solutions of the Bethe ansatz equations, the set of roots
$\{z_1,z_2,\ldots,z_N\}$  to be used in eq.(\ref{W}) is chosen to
be the set for which the energy parameter of the corresponding state is
the lowest (usually it is the ground state).  The required electric
field which gives rise to the $N+1$ solvable states is then
obtained as
\be
\mu E^{(N)}_r=W_N + |k|/r ~. \label{E-N} \ee

For the spherically symmetric electric fields which we consider in
this section, there are three possible types of $sl(2)$-based QES
field configurations.  They belong to Class VII, VIII, and IX in
the classification listed in \cite{Tur}. Below we shall illustrate
our construction of QES electric fields through Class VII QES
systems.

\vskip 0.8cm

\centerline{{\bf Class VII}}

The general potential in Class VII has the form
\be
V_N(r)=a^2r^6+2abr^4+\left[b^2-a\left(4N+2\gamma
+3\right)\right]r^2
+\gamma\left(\gamma-1\right)r^{-2}-b\left(2\gamma+1\right)~,
\label{cVII}
\ee
where $a,b$ and $\gamma$ are constants.
The gauge function is
\be
g(r)=\frac{a}{4}r^4 + \frac{b}{2}r^2 -\gamma\ln {r}~. \label{g4}
\ee We must have $a,\gamma >0$ to ensure normalizability of the
wave function.  Eqs.(\ref{g4}) and (\ref{W2}), together with the
relation $W_0(r)=g^\prime(r)$, give us the electric field
$E_r^{(0)}$:
\be
\mu E_r^{(0)} (r)= ar^3 + b r~. \ee The Dirac-Pauli equation with
this field configuration admits a QES ground state with energy
${\cal E}^2=m^2$ ($\epsilon=0$) and ground state function
$f_-\propto \exp(-g_0(r))$. Also, here we have $\gamma=|k|$. We
retain the symbol $\gamma$ so that some general formulae in this
section can be carried over in the next section simply with
$\gamma$ redefined.

To determine electric field configurations admitting QES
potentials $V_N$ with higher weight, we need to obtain the Bethe
ansatz equations for $\phi$.
Letting $z=h(r)=r^2$, eq.(\ref{phi}) becomes
\begin{eqnarray}
\left[-4 z \frac{d^2}{dz^2} +\left(4az^2 +4bz -2\left(2\gamma
+1\right) \right)\frac{d}{dz} -\left(4aNz + \epsilon
\right)\right]\phi(z) =0~. \label{phi-VII}
\end{eqnarray}
In terms of the $sl(2)$ generators $J^+,~J^-$ and $J^0$, the
differential operator acting on $\phi(z)$ in eq.(\ref{phi-VII}) can be
written as \begin{eqnarray}
T_{VII}=-4 J^0 J^- + 4a J^+ +4bJ^0 - 2\left(N+2\gamma +1\right)J^-
+ \rm{constant}~. \label{T-VII}
\end{eqnarray}
For $N=0$, the value of the $\epsilon$ is $\epsilon=0$. For higher
$N>0$ and $\phi(r)=\prod_{i=1}^N (z-z_i)$, the electric field
$E_r^{(N)}(r)$ is obtained from eqs.(\ref{W}) and (\ref{E-N}):
\begin{eqnarray}
\mu E_r^{(N)}(r) = \mu E_r^{(0)}(r) - \sum_{i=1}^N
\frac{h^\prime(r)}{h(r)-z_i}~. \label{EN}
\end{eqnarray}
For the present case, the roots $z_i$'s are found from the Bethe
ansatz equations
\begin{eqnarray}
2az_i^2 +2bz_i -\left(2\gamma+1\right) - \sum_{l\neq
i}\frac{z_i}{z_i-z_l} =0~, \quad\quad i=1,\ldots,N~,
\label{BA-VII}
\end{eqnarray}
and $\epsilon$ in terms of the roots $z_i$'s is
\begin{eqnarray}
\epsilon=2\left(2\gamma+1\right)\sum_{i=1}^N \frac{1}{z_i}~.
\label{E-VII}
\end{eqnarray}

For $N=1$ the roots $z_1$ are
\begin{eqnarray}
z_1^\pm=\frac{-b\pm\sqrt{b^2+2a(2\gamma+1)}}{2a}~, \label{z1}
\end{eqnarray}
and the values of $\epsilon$ are
\begin{eqnarray}
\epsilon^\pm=2\left(b\pm\sqrt{b^2+2a(2\gamma+1)}\right)~.
\end{eqnarray}
 For $a>0$, the root $z_1^-=-|z_1^-|<0$ gives the
ground state. With this root, one gets the superpotential
\begin{eqnarray}
W_1(r)=ar^3 +br -\frac{2r}{r^2+|z_1^-|} -\frac{\gamma}{r}~.
\end{eqnarray}
From eq.(\ref{EN}), the corresponding electric field is
\be
\mu E_r^{(1)} (r) &=& ar^3 +br - \frac{2r}{r^2+ |z_1^-|}~.
\ee
The QES potential appropriate for the problem is
\begin{eqnarray}
V(x) &=& W_1^2-W_1^\prime~,\cr &=&V_1 -\epsilon~.
\end{eqnarray}
The one-dimensional SUSY always sets the energy parameter of
ground state at $\epsilon=0$.  Hence, the ground state and the
excited state have energy parameter $\epsilon=0$ and
$\epsilon=\epsilon^+ -\epsilon^-=4\sqrt{b^2+2a(2\gamma+1)}$, and
wave function
\be
f_-\propto e^{-g_0(r)}\left(r^2-z_1^-\right) \ee and
\be
f_-\propto e^{-g_0(r)}\left(r^2-z_1^+\right)~, \ee
respectively.

QES potentials and electric fields for higher degree $N$ can be constructed
in the same manner.

\vskip 1 truecm

\centerline{\mbox{\large{\bf IV.  Electric fields in cylindrical
coordinates}}}

We now treat the Dirac-Pauli equation in the cylindrical
coordinates.  In this case, it is easier to solve $H^2\Phi={\cal
E}^2 \Phi$ instead of $H\psi={\cal E} \psi$. Once $\Phi$ is
solved, the desired solution $\psi$ is given by $\psi= (H + {\cal
E})\Phi$. Setting $\Phi=(f_-,f_+)^t$, the equation $(H^2-{\cal
E}^2)\Phi=0$ becomes
\be
\left[p^2\mp \mu\nabla\cdot {\bf E} + \mu^2 E^2 \pm i\mu {\bsigma}\cdot
\left(2 {\bf E}\times \nabla - \nabla\times {\bf
E}\right)\right]f_\mp=\left({\cal E}^2 - m^2\right)f_\mp~.
\label{eqn-f}
\ee

We consider electric field configuration which has cylindrical symmetry,
namely,
\be
E_x=xf(\rho)~, ~~E_y=yf(\rho)~, ~~E_z=0~, \ee where
$\rho=\sqrt{x^2+y^2}$, and $f(\rho)$ is some function of $\rho$.
It can be checked that the conserved quantities are $p_z$ and
$J_z=L_z+\Sigma_3/2$. Thus the eigenstates can be chosen as
\be
f_-=\frac{e^{ik_zz}}{\sqrt{\rho}}\left(\begin{array}{c} R_1(\rho)e^{il\phi}\\
R_2(\rho)e^{i(l+1)\phi}
\end{array}\right)~~,~~~
f_+=\frac{e^{ik_zz}}{\sqrt{\rho}}\left(\begin{array}{c} R_3(\rho)e^{il\phi}\\
R_4(\rho)e^{i(l+1)\phi}
\end{array}\right)~.
\ee $\Phi$ is an eigenstate of $p_z$ and $J_z$ with eigenvalues
$k_z$ ($k_z$ real) and $j_z=l+1/2$ ($l=0,\pm 1,\pm 2\ldots$),
respectively. Eq.(\ref{eqn-f}) reduce to four decoupled ones:
\be
\left[-\frac{d^2}{d\rho^2}+\mu\rho^2f^2 - 2\mu
f\left(l+1\right)-\mu\rho f^\prime +
\frac{l^2-\frac{1}{4}}{\rho^2}\right] R_1(\rho) &=& \epsilon
R_1(\rho)~,\label{R1}\\ \left[-\frac{d^2}{d\rho^2}+\mu\rho^2f^2 +
2\mu f l -\mu\rho f^\prime +
\frac{(l+1)^2-\frac{1}{4}}{\rho^2}\right] R_2(\rho) &=& \epsilon
R_2(\rho)~,\label{R2}\\ \left[-\frac{d^2}{d\rho^2}+\mu\rho^2f^2 +
2\mu f\left(l+1\right)+\mu\rho f^\prime +
\frac{l^2-\frac{1}{4}}{\rho^2}\right] R_3(\rho) &=& \epsilon
R_3(\rho)~,\label{R3}\\ \left[-\frac{d^2}{d\rho^2}+\mu\rho^2f^2 -
2\mu f l + \mu\rho f^\prime +
\frac{(l+1)^2-\frac{1}{4}}{\rho^2}\right] R_4(\rho) &=& \epsilon
R_4(\rho)~,\label{R4} \ee where $\epsilon\equiv {\cal
E}^2-m^2-k_z^2$. From the above equations it follows that
eqs.(\ref{R1}) and (\ref{R4}) represent a pair of one-dimensional
SUSY partners \cite{HoRoy}, with superpotential given by
\be
W(\rho)=\mu \rho f(\rho)-\frac{\gamma}{\rho}~, ~~~\gamma=|l|+1/2~.
\label{W3} \ee Eqs.(\ref{R2}) and (\ref{R3}) form another pair of SUSY partners,
obtainable from eqs.(\ref{R4}) and (\ref{R1}), respectively, by
changing $\mu$ to $-\mu$. Results in \cite{HoRoy} can therefore be
carried over directly with minor modifications.

To be specific, let us assume $\mu f>0$. For $l\geq 0$, we have
$R_2=R_3=0$.  The ground state is given by
\be
R_1&\propto& \exp(-g(\rho))~,\\
 g(\rho)&=&\mu\int^\rho \rho^\prime
f(\rho^\prime) d\rho^\prime -\gamma\ln \rho~, \ee and $R_4=0$. For
excited states, the components $R_1$ and $R_4$ are related by
$R_4\propto A^+ R_1$, where $A^+$ is defined by eq.(\ref{A}) and
(\ref{W3}). For $l\leq -1$, we simply interchange $R_1$ and $R_2$,
and  $R_3$ and $R_4$.

Comparing eq.(\ref{W3}) with (\ref{W2}), it is seen that all
results in Sect. III can be carried over by changing $r$ , $E_r$
and $\gamma=|k|$ in eq.(\ref{W2}) to $\rho$, $\rho f$, and
$\gamma=|l|+1/2$, respectively. Hence we will not discuss this
case in details, but only quote some main results.  For instance,
the exactly solvable cases are:

~~~~~~~~~~i)   oscillator-like :  $\mu f(\rho)\propto {\rm
constant} $~;

~~~~~~~~~~ii)  Coulomb potential-like :   $\mu f(\rho)\propto
1/\rho $~;

~~~~~~~~~~iii) zero field-like :  $\mu f(\rho)\propto 1/\rho^2 $~.

As in Sect III, the quasi-exactly solvable cases are Classes VII,
VIII and IX of QES systems based on the $sl(2)$ algebra as
classified by Turbiner \cite{Tur}.  For the case of Class VII, the
simplest two configurations of the electric fields are obtained
from the function $f(\rho)$ as follows:

~~~~~~i) one-state case:    $\mu f^{(0)}(\rho) = a \rho^2 + b ~
(a>0)~;$

~~~~~~ii) two-state case:   $\mu f^{(1)}(\rho) = a \rho^2 + b  -
2/(\rho^2 + |z_1^-|)~,$ where $z_1^-$ is given by eq.(\ref{z1}).

\vskip 1 truecm

 \centerline{\mbox{\large{\bf V.  Electric fields
in Cartesian coordinates}}}

Finally we consider electric field configurations in one
dimension.  For definiteness, we assume
\be
E_x(x)=\bar{W}(x)~,~~ E_y=E_z=0~.
\label{E}
\ee
The conserved quantities in this case are $p_y$ and $p_z$, with eigenvalues
$k_y$ and $k_z$, respectively.
Eigenstate can be written as
\be
\psi = e^{ik_yy+ ik_zz} \left(\begin{array}{c} f_-(x)\\ f_+(x)
\end{array}\right)~.
\label{psi} \ee Substituting eqs.(\ref{E}) and (\ref{psi}) into
(\ref{eqn-f}), we get
\be
\left[-\frac{d^2}{dx^2} + \left(\mu \bar{W}(x)\mp \sigma_3 k_y\right)^2
\mp \mu \bar{W}^\prime\right]~f_\mp =\left({\cal E}^2-m^2-k_z^2\right)~
f_\mp~. \label{lin} \ee Again, these equations are in the same
form as eq.(3) in \cite{HoRoy} and eq.(408) in \cite{Cooper}.
Similar to the case in Section IV, the four components of $\psi$ are
related by two one-dimensional SUSY, namely, the 1st (2nd)
component of $f_-$ and the 2nd (1st) component of $f_+$ form a
one-dimensional SUSY pair. As such, this system can be dealt with
according to the general procedures outlined in the last section
and in \cite{HoRoy}. Here we shall only give the classifications
of the electric field configurations which permit exact and
quasi-exact solutions.

\vskip 0.8cm
 \centerline{{\bf A.  Exactly solvable cases}} The
exactly solvable cases have been classified in \cite{Cooper}.
There are three types of exactly solvable field configurations,
namely,
\be
\mu \bar{W}\mp k_y = \left\{ \begin{array}{ll}
               \alpha x+c &\mbox{(shifted oscillator-like)~;}\\
                c_1 \exp(-\alpha x) +c_2 & \mbox{(Morse-potential)~;}\\
               \tanh(\alpha x) + c & \mbox{(Rosen-Morse II)~;}
               \end{array}
               \right.
\ee
Here $c,c_1,c_2$ and $\alpha$ are constants.

\vskip 0.8cm
\centerline{{\bf B.  Quasi-exactly solvable cases}}

From the discussions given in \cite{HoRoy}, the system of
eq.(\ref{lin}) are QES based on the $sl(2)$ algebra.  In fact, six
classes of QES electric field configurations can be constructed.
These classes correspond to Class I to Class VI in Turbiner's
classification.  One can follow the general procedures described
in Sect. III to construct in each class the electric fields which
admit different numbers of QES states, including excited states.
Here we only present the main results for Class I, referring the
readers to \cite{HoRoy} for further details.

\vskip 0.8cm \centerline{\bf Class I}

We concentrate on the first set of one-dimensional SUSY system in
eq.(\ref{lin}),
namely, the first component of $f_-$ and the second component of $f_+$.
The superpotential for the system is
\be
W(x)\equiv \mu \bar{W}(x)-k_y~.
\ee

According to Turbiner's classification, the QES potential
belonging to Class I has the form
\begin{eqnarray}
V_N(x)=a^2 e^{-2\alpha x}-
a\left[\alpha(2N+1)+2b\right]e^{-\alpha x}+
c\left(2b-\alpha\right)e^{\alpha x} + c^2 e^{2\alpha
x}+b^2-2ac~. \label{V-I}
\end{eqnarray}
Without loss of
generality, we assume $\alpha, a,c>0$ for definiteness.  The
corresponding gauge function $g(x)$ is given by
\begin{eqnarray}
g(x)= \frac{a}{\alpha} e^{-\alpha x} + \frac{c}{\alpha} e^{\alpha
x} + b x~. \label{g-I}
\end{eqnarray}
One should always keep in mind that the parameters selected must
ensure convergence of the function $\exp(-g(x))$ in order to
guarantee normalizability of the wave function .  The potential
$V(x)$ that gives the ground state, with energy parameter
$\epsilon\equiv {\cal E}^2-m^2-k_z^2=0$, is generated by
\begin{eqnarray}
V(x)&=&V_0 \cr &=& W_0^2-W_0^\prime~,
\end{eqnarray}
with
\begin{eqnarray}
 W_0(x)&=&g^\prime(x)\cr &=&-a e^{-\alpha x}+ c e^{\alpha
x}+b~.
\end{eqnarray}
The corresponding electric field is
\begin{eqnarray}
\mu E_x^{(0)} = -a\alpha e^{-\alpha x} + c\alpha
e^{\alpha x}~+b+k_y~.
\end{eqnarray}

To obtain electric fields and the corresponding potentials which
admit solvable states with higher weights $j$, we must first
derive the Bethe ansatz equations. To this end, let us perform the
change of variable $z=h(x)=\exp(-\alpha x)$. Eq.(\ref{phi}) then
becomes
\begin{eqnarray}
\left\{-\alpha z^2 \frac{d^2}{dz^2} +\left[2az^2 -(2b+\alpha)z
-2c\right]\frac{d}{dz} +\left[-2aNz -
\frac{\epsilon}{\alpha}\right]\right\}\phi(z) =0~. \label{phi-I}
\end{eqnarray}
The differential operator acting on $\phi(z)$ can be written as a
quadratic combination of the $sl(2)$ generators $J^+,~J^-$ and $J^0$ as
\begin{eqnarray}
T_I=-\alpha J^+ J^- + 2a J^+ -
\left[\alpha(N+1)+2b\right]J^0 -2cJ^- + \rm{constant}~.
\label{T-I}
\end{eqnarray}

For $N>0$, there are $N+1$ solutions which include excited states.
Assuming $\phi(z)=\prod_{i=1}^N (z-z_i)$ in eq.(\ref{phi-I}), one
obtains the Bethe ansatz equations which determine the roots
$z_i$'s
\begin{eqnarray}
2az_i^2 -(2b+\alpha)z_i - 2c -2\alpha \sum_{l\neq
i}\frac{z_i^2}{z_i-z_l} =0~, \quad\quad i=1,\ldots,N~,
\label{BA-I}
\end{eqnarray}
and the equation which gives the energy parameter in terms of the roots
$z_i$'s
\begin{eqnarray}
\epsilon =2\alpha c\sum_{i=1}^N \frac{1}{z_i} ~. \label{E-I}
\end{eqnarray}
Each set of \{$z_i$\} determine a QES energy $E$ with the
corresponding polynomial $\phi$.

As an example, consider the $j=1/2$ case with $N=1$ and
$\phi(z)=z-z_1$.  There are two solutions.  From eq.(\ref{BA-I}),
one sees that the root $z_1$ satisfies
\begin{eqnarray}
2az_1^2 -(2b+\alpha)z_1 -2c =0~,
\end{eqnarray}
which gives two solutions
\begin{eqnarray}
z_1^\pm=\frac{(2b+\alpha)\pm\sqrt{(2b+\alpha)^2 + 16ac}}{4a}~.
\end{eqnarray}
The corresponding energy parameters are
\begin{eqnarray}
\epsilon^\pm&=&  2\alpha c\frac{1}{z_1}\cr &=&
-\frac{\alpha}{2}\left[(2b+\alpha)\mp\sqrt{(2b+\alpha)^2 +
16ac}\right]~.
\end{eqnarray}
For the parameters assumed here, the solution with root
$z_1^-=-|z_1^-|<0$ gives the ground state, while that with root
$z_1^+>0$ gives the first excited state. The superpotential $W_1$
is constructed according to eq.(\ref{W})
\begin{eqnarray}
W_1(x)&=&W_0-\frac{h^\prime(x)}{h(x)-z_1^-}\cr &=&-ae^{-\alpha x}+
c e^{\alpha x}+\frac{\alpha}{1+|z_1^-|e^{\alpha x}}+b~.
\end{eqnarray}
This gives the electric field
\begin{eqnarray}
\mu E_x^{(1)}(x)=-ae^{-\alpha x}+
c e^{\alpha x}+\frac{\alpha}{1+|z_1^-|e^{\alpha x}}+b+k_y~.
\end{eqnarray}
The ground state and the excited state have
energy ${\cal E}^2=m^2+k_z^2$ and ${\cal E}^2=m^2+k_z^2+ \alpha
\sqrt{(2b+\alpha)^2
+ 16ac}$, respectively.

\bigskip
\centerline{\mbox{\large \bf VI. Conclusions}}

In this paper we have shown that the Dirac-Pauli equation
describing a neutral particle with anomalous magnetic moment in an
external electric field is a QES system.  Forms of electric field
configurations permitting exact solutions, and QES states based on
$sl(2)$  algebra are classified in the spherical, cylindrical, and
Cartesian coordinates.

We have also demonstrated that the Dirac-Pauli equation with only
electric fields are equivalent to generalized Dirac oscillators.
Thus all the results obtained here are directly applicable to the
later.  Particularly, the exact solvability of the original Dirac
oscillator is now recognized as one of the three exactly solvable
cases of the Dirac-Pauli equation.

It has been shown that the two-dimensional Dirac-Pauli equation is
equivalent to a two-dimensional Dirac equation minimally  coupled
to a vector potential by a duality transformation
\cite{AC-eff,lin2}. However, as mentioned in Sect.I,  the square
of the Dirac Hamiltonian minimally coupled to a vector potential
is proportional to the Pauli equation up to an additive constant.
Hence the exact solvability discussed in \cite{Cooper} and the
quasi-exactly solvability discussed in \cite{HoRoy} of the Pauli
equation can be straightforwardly applied to the two-dimensional
Dirac-Pauli equation.

\vskip 2cm \centerline{\bf Acknowledgments}

This work was supported in part by the Republic of China through
Grant No. NSC 92-2112-M-032-015.
 P.R. would like to thank the
Department of Physics at Tamkang University for support during his
visit.

\end{document}